\documentclass{myaa}
\usepackage{graphicx}
\usepackage{txfonts}
\usepackage{natbib}
\usepackage[figuresright]{rotating}
\usepackage[a4paper,breaklinks]{hyperref}

\bibpunct{(}{)}{;}{a}{}{,} 

\begin{document}
\def\teff{$T\rm_{eff}$ }
\def\kms {\,$\mathrm{km\,s^{-1}}$}
\def\ms {$\mathrm{m\, s^{-1}}$ }

\newcommand{\Teff}{\ensuremath{T_\mathrm{eff}}}
\newcommand{\gf}{\ensuremath{gf}}
\newcommand{\g}{\ensuremath{g}}
\newcommand{\loggf}{\ensuremath{\log\gf}}
\newcommand{\glog}{\ensuremath{\log\g}}
\newcommand{\logg}{\ensuremath{\log\,g}}
\newcommand{\pun}[1]{\,#1}
\newcommand{\cobold}{\ensuremath{\mathrm{CO}^5\mathrm{BOLD}}}
\newcommand{\linfor}{Linfor3D}
\newcommand{\punms}{\mbox{\rm\,m\,s$^{-1}$}}
\newcommand{\punkms}{\mbox{\rm\,km\,s$^{-1}$}}
\newcommand{\abuhe}{\mbox{Y}}
\newcommand{\grav}{\ensuremath{g}}
\newcommand{\mlp}{\ensuremath{\alpha_{\mathrm{MLT}}}}
\newcommand{\mlpcm}{\ensuremath{\alpha_{\mathrm{CMT}}}}
\newcommand{\moh}{\ensuremath{\left[\mathrm{M/H}\right]}}
\newcommand{\senv}{\ensuremath{\mathrm{s}_{\mathrm{env}}}}
\newcommand{\shelio}{\ensuremath{\mathrm{s}_{\mathrm{helio}}}}
\newcommand{\smin}{\ensuremath{\mathrm{s}_{\mathrm{min}}}}
\newcommand{\spun}{\ensuremath{\mathrm{s}_0}}
\newcommand{\sstar}{\ensuremath{\mathrm{s}^\ast}}
\newcommand{\tauross}{\ensuremath{\tau_{\mathrm{ross}}}}
\newcommand{\ttaurelation}{\mbox{T$(\tau$)-relation}}
\newcommand{\Ysurf}{\ensuremath{\mathrm{Y}_{\mathrm{surf}}}}

\newcommand{\draftflag}{false}

\newcommand{\beq}{\begin{equation}}
\newcommand{\eeq}{\end{equation}}
\newcommand{\pdx}[2]{\frac{\partial #1}{\partial #2}}
\newcommand{\pdf}[2]{\frac{\partial}{\partial #2}\left( #1 \right)}

\newcommand{\var}[1]{{\ensuremath{\sigma^2_{#1}}}}
\newcommand{\sig}[1]{{\ensuremath{\sigma_{#1}}}}
\newcommand{\cov}[2]{{\ensuremath{\mathrm{C}\left[#1,#2\right]}}}
\newcommand{\xtmean}[1]{\ensuremath{\left\langle #1\right\rangle}}

\newcommand{\mD}{\ensuremath{\left\langle\mathrm{3D}\right\rangle}}

\newcommand{\eref}[1]{\mbox{(\ref{#1})}}

\newcommand{\Vact}{\ensuremath{\nabla}}
\newcommand{\Vad}{\ensuremath{\nabla_{\mathrm{ad}}}}
\newcommand{\Veddy}{\ensuremath{\nabla_{\mathrm{e}}}}
\newcommand{\Vrad}{\ensuremath{\nabla_{\mathrm{rad}}}}
\newcommand{\Vraddiff}{\ensuremath{\nabla_{\mathrm{rad,diff}}}}
\newcommand{\cp}{\ensuremath{c_{\mathrm{p}}}}
\newcommand{\taueddy}{\ensuremath{\tau_{\mathrm{e}}}}
\newcommand{\vconv}{\ensuremath{v_{\mathrm{c}}}}
\newcommand{\Fconv}{\ensuremath{F_{\mathrm{c}}}}
\newcommand{\lmix}{\ensuremath{\Lambda}}
\newcommand{\Hp}{\ensuremath{H_{\mathrm{P}}}}
\newcommand{\Hptop}{\ensuremath{H_{\mathrm{P,top}}}}
\newcommand{\COBOLD}{{\sc CO$^5$BOLD}}

\newcommand{\changed}{}

\newcommand{\I}{\ensuremath{I}}
\newcommand{\Irot}{\ensuremath{\tilde{I}}}
\newcommand{\F}{\ensuremath{F}}
\newcommand{\Frot}{\ensuremath{\tilde{F}}}
\newcommand{\vsini}{\ensuremath{V\sin(i)}}
\newcommand{\vvsini}{\ensuremath{V^2\sin^2(i)}}
\newcommand{\vsinimu}{\ensuremath{\tilde{v}}}
\newcommand{\rotint}{\ensuremath{\int^{+\vsinimu}_{-\vsinimu}\!\!d\xi\,}}
\newcommand{\imu}{\ensuremath{m}}
\newcommand{\imupone}{\ensuremath{{m+1}}}
\newcommand{\nmu}{\ensuremath{N_\mu}}
\newcommand{\msum}[1]{\ensuremath{\sum_{#1=1}^{\nmu}}}
\newcommand{\wmu}{\ensuremath{w_\imu}}

\title{Sulphur abundances from the \ion{S}{i} near-infrared triplet at 1045\pun{nm}
\thanks{based on data from the UVES Paranal Observatory Project (ESO DDT 
Program ID 266.D-5655)}
}

\author{
E. Caffau     \inst{1}\and
R. Faraggiana \inst{2}\and
P. Bonifacio  \inst{3,1,4}\and
H.-G. Ludwig  \inst{3,1}\and
M. Steffen    \inst{5}
}
\institute{GEPI, Observatoire de Paris, CNRS, Universit\'e Paris Diderot ; Place 
Jules Janssen 92190
Meudon, France
\and
Universit\`a degli Studi di Trieste, Dipartimento di Astronomia,
Via Tiepolo 11, I-34131 Trieste, Italy
\and
CIFIST Marie Curie Excellence Team
\and
Istituto Nazionale di Astrofisica - Osservatorio Astronomico di
Trieste, Via Tiepolo 11, I-34131 Trieste, Italy
\and
Astrophysikalisches Institut Potsdam, An der Sternwarte 16, D-14482 Potsdam, Germany
}
\authorrunning{Caffau et al.}
\titlerunning{Sulphur abundances from the \ion{S}{i} near-infrared triplet at 1045\pun{nm}}
\offprints{Elisabetta.Caffau@obspm.fr}
\date{Received ...; Accepted ...}

\abstract
{Unlike silicon  and calcium, sulphur is an $\alpha$-element 
which does not form dust. 
Some of the available observations of the evolution of sulphur with
metallicity indicate an increased scatter of
sulphur to iron ratios at low metallicities or
even a bimodal distribution, with some stars showing
constant S/Fe at all metallicities and others showing 
an increasing S/Fe ratio with decreasing metallicity. 
In metal-poor stars \ion{S}{i} lines of Multiplet~1 at 920\,nm
are not
yet too weak to permit to measure the sulphur abundance~A(S), however in ground-based
observations they are severely affected by telluric
lines.}
{We investigate the possibility to measure sulphur abundances from 
\ion{S}{i} Mult.~3 at 1045\,nm
lines, which lie in the near infra-red. 
These are slightly weaker  than those of Mult.~1, but lie in a
range not affected by telluric lines.}
{We investigate the lines of Mult.~3
in the  Sun (G2V), Procyon (F5V), HD~33256 (F5V), HD~25069 (G9V) and 
$\epsilon$ Eri (HD~22049, K2V).
For the Sun and Procyon 
the analysis has been performed with \cobold\
3D hydrodynamical model atmospheres, 
for the other three stars, for which hydrodynamical simulations are not
available, the analysis has been performed using 1D model atmospheres.
}
{For our sample of stars we find a global agreement of A(S) from lines 
of different multiplets.
}
{Our results suggest that 
the infrared lines of Mult.~3 are a viable indicator
of the sulphur abundance which, because of
the intrinsic strength of this multiplet,  should  be suitable
to study the trend of [S/Fe] at low metallicities.
}

\maketitle


\section{Introduction}

The so-called $\alpha$-elements (O, Ne, Mg, Si, S, Ca) are among the
main products of Type II Supernovae
\citep{1995ApJS..101..181W,2003ApJ...592..404L,2004ApJ...608..405C}. 
The iron peak elements are
produced by Type II SNe,
and it is commonly accepted that 
Type Ia
Supernovae produce similar, or larger amounts
of iron-peak elements, and produce little or no $\alpha$-elements,
\citep{1984ApJ...279L..23N,1999ApJS..125..439I}.
The different timescales for the first explosions of Type II or Type Ia
SNe to occur, makes the abundance ratio of $\alpha$-elements to  iron-peak elements
a powerful diagnostics of the chemical evolution and star
formation history of   a galaxy. 
In the Milky Way, stars of lower metallicity
are characterised by higher $\alpha$-to-iron abundance ratios
than found in the Sun and solar metallicity stars
\citep[see, for example,][]{1988A&A...191..121B,2003A&A...406..131G,2004A&A...416.1117C}. This is
usually interpreted in terms of a lower contribution
of Type Ia SNe.
Systems which are characterised by low or bursting star formation, 
like dwarf galaxies, give time to Type Ia SNe to explode
before the enrichment due to Type II SNe has greatly increased.
Consequently, such systems display rather low $\alpha$-to-iron ratios
even at low metallicities \citep[see][ and references therein]{2004AJ....128.1177V}
and at solar metallicities, display sub-solar ratios
\citep{2004A&A...414..503B,2005A&A...441..141M}.

One should be aware that the above outlined,
simple interpretation of the  $\alpha$-to-iron peak
element ratios in terms   of products of Type II and
Type Ia SNe, relies on the nucleosynthesis
computed with  1D explosion models. For Type Ia
SNe such computations predict that
the original C-O white dwarf is totally burned, mainly
to $^{56}$Ni. It is interesting to note that in the 
2D models by \citet{2005NuPhA.758..451B}, most of the
white dwarf is not burnt and roughly equal masses of
$^{56}$Ni and $^{28}$Si are produced.
It is clear that a different explanation of the $\alpha$-element
enhancement must be sought if these computations are confirmed.

While  the theoretical interpretation of
the $\alpha$-to-iron peak ratios is likely
still open to debate, it is clear that, from 
the observational point of view,  
the abundance of $\alpha$-elements is an important 
property of any stellar population.
For the study of chemical
evolution in external galaxies,
the more readily available objects 
are Blue Compact Galaxies (BCGs) through analysis
of their emission line spectra, and Damped Ly-$\alpha$
systems (DLAs)
through the analysis of resonance absorption
lines. In this way, it is relatively easy to measure
sulphur in the gaseous component of both groups of galaxies
\citep{1989ApJ...345..282G,2000ApJ...536..540C}.

At variance with  Si and Ca, S is a volatile element,
and therefore it is not locked into dust grains in the interstellar medium,
so that  no correction
is needed to the measured sulphur abundance.
This makes sulphur a more convenient element to trace
the $\alpha$'s than either Si or Ca;
all three elements are  made by 
oxygen burning, either in a central burning
phase, convective shell, or explosive phase. According to
\citet{2003MSAIS...3...58L}, 
there is thus a strong reason to believe that Si, S, and Ca
vary in lockstep during the chemical evolution.
In spite of this, it is
certainly unsatisfactory to compare
S/Fe in external galaxies to Si/Fe or Ca/Fe
in the Milky Way. It is much more convenient
to use a 
reliable Galactic reference
for sulphur abundances, which may be directly
compared to measures in  external galaxies.

The only way 
to measure  abundances at different metallicities
in the Galaxy is to use stars.
Unfortunately there are very few sulphur lines which 
are not blended and remain strong enough to be measured
at low metallicities.
The lines of Mult.~8\footnote{We use the Multiplet numbering by \citet{1945CoPri..20....1M}}
(675\pun{nm}) and~6 (869\pun{nm}) 
are weak, so only detectable in
solar or moderately metal-poor stars, 
hardly lower
than [Fe/H] $\sim$ --1.5 (for the Mult.~8), 
or --2.0 (for Mult.~6).
The lines of Mult.~1 (920\pun{nm}) have been recently used to measure 
A(S)\footnote{A(S)=log (N(S)/N(H)) + 12} because
these lines are strong, and are detectable even at low metallicity.
The non-negligible  NLTE effects in these lines
make them even stronger \citep{2005PASJ...57..751T},
therefore more easily measurable.
The main problem is that the range in wavelength in which
Mult.~1 lies is contaminated by numerous
telluric lines. This makes it difficult to obtain all (or at least one)
components of Mult.~1 unaffected by telluric absorption.

The 1045\pun{nm} lines of Mult.~3 are particularly suited to measure the
sulphur abundance. Even if the lines are not as strong as the
components of Mult.~1, the big advantage is that there are no
telluric lines present in the vicinity of their wavelength.
Observing sulphur lines of Mult.~3 provides a possibility
to obtain a reliable sulphur abundance in very metal-poor stars.

Existing studies of sulphur in Galactic stars have suggested that
in the range $-2.5<$[Fe/H]$<-2.0$ the [S/Fe] ratio
shows either a large scatter or a bimodal behaviour;
Most of the stars converge to a ``plateau'' at about [S/Fe]=+0.4, while
a  non-negligible number of stars  shows a ``high'' value
of [S/Fe] around +0.8.
In the sample of \citet{2005A&A...441..533C}, the determination
of [S/Fe] in this range of metallicity is based on the
non contaminated lines of Mult.~1.
This behaviour has no proposed theoretical interpretation and
it may well be due to systematic errors related to
the use of Mult.~1. 
It is therefore of great interest to verify this puzzling
finding by the use of an independent and, hopefully better,
diagnostics of the sulphur abundance, as can be afforded
by the lines of Mult.~3.
We moreover recall that these are the only strong \ion{S}{i}
lines belonging to a triplet system instead of quintet systems as
the other \ion{S}{i} lines in the visual and near-IR range.

The aim of the paper is to
study the  \ion{S}{i} lines of Mult.~3
in  two well known
stars -- the Sun and Procyon -- and to investigate three
other bright field stars of spectral type F, G and K: 
HD~22049, HD~33256, HD~25069 respectively.
We compare A(S) from two sulphur lines widely used
in sulphur abundance determinations, 
to the one from Mult.~3 lines.
We want to establish Mult.~3 as valuable abundance indicator.


\section{Atomic data}

The \ion{S}{i} lines that we consider in this work, are reported
in Table \ref{atomicdata}. The triplet at 675.7\pun{nm} and 
the line at 869.4\pun{nm} have been widely used in the determination
of sulphur abundances  
\citep[see for instance,][]{2004AA...415..993N,2005A&A...441..533C}.
The lines of Mult.~3 have not yet been considered\footnote{%
The referee made us aware of a
  paper (Nissen et al. 2007 astro-ph/0702689, submitted to A\&A) which became
  available through arXiv (\href{arxiv.org}{arxiv.org}) after the submission
  of this work. The paper describes the first use of the 1045\pun{nm}
  lines for abundance work in metal-poor stars.},
except in the Sun \citep{1978MNRAS.183...79L,2005PASJ...57..751T}.
We know that the bluest line of Mult.~3 is blended with
an iron line of poorly known \loggf, but we nevertheless
keep this line
for the Sun and Procyon for which we have good observed spectra in hand,
and for HD~33256 where A(Fe) is lower
than in the Sun, as well as for $\epsilon$ Eri 
where  the iron contribution is only 12\,\% of the total equivalent width (EW);
we discard this line for HD~25069 for which
the contribution of iron to  the total EW is 
about 18\,\%, comparable to the error
of the EW for this star.
In any case, the relative
contribution of the iron line to the blend becomes
smaller for very metal-poor stars, and thus
the line can be a good indicator of the sulphur abundance.

The \loggf~value of the Mult.~3 lines, as those of all the other \ion{S}{i}
lines used here, have been taken from the Kurucz line list;
the data are approximately coincident with those from the 
NIST Atomic Spectra Database 
(\href{http://physics.nist.gov/PhysRefData/ADS}{http://physics.nist.gov/PhysRefData/ADS}).
These \gf ~ are experimental values based on 
three experiments. Owing to the moderate accuracy of the absolute
scale, \citet{1969atp..book.....W} renormalised the data theoretically and
judged the resulting accuracy of the \gf\ to be within 50\,\%.
In particular we note that all the lines selected for S abundance 
determination are
of D quality, i.e. the uncertainty of the oscillator strength
is \mbox{$\le 50$\,\%}; the lines with highest accuracy 
(D+, i.e. an uncertainty of the oscillator strength $\le 40$\,\%) 
are those of Mult.~1
(920\pun{nm}), Mult.~3 (1045\pun{nm}), and the line at 869.4\pun{nm} 
of Mult.~6, while the line at 869.3\pun{nm} of Mult.~6 is 
of D quality, as is one component (675.7171\pun{nm}) of the 
675.7\pun{nm} line of Mult.~8.  One component of the 675\pun{nm} 
triplet of Mult.~8, the 675.6851\pun{nm}
line, is of E quality 
(uncertainty of the oscillator strength $> 50$\,\%).

\begin{table}
\caption{Atomic parameters of the sulphur lines.
Col. (1) is the wavelength; col. (2) the 
multiplet number; col. (3) the transition; col. (4)
the \loggf ~ of the transition taken from the Kurucz line list; 
col. (5) the excitation energy;
col. (6) the Van der Waals damping constant computed at a
temperature of 5\,500\pun{K} according to ABO theory (see
Sect.~\ref{s:dataanalysis}) or Kurucz 
approximation (see text).}
\label{atomicdata}
\begin{center}
\begin{tabular}{rrrrrr}
\hline
\noalign{\smallskip}
Wavelength& Mult. & Transition               & \loggf  & $\chi _{\rm lo}$ & $\log \gamma_6/{\rm N_{\rm H}}$ \\
(nm) air  &       &                          &         &              (eV)&  $\left(\mathrm{s}^{-1}~\mathrm{cm}^3\right)$\\
 (1)      & (2)   &    (3)                   & (4)     & (5)              & (6)\\
\noalign{\smallskip}
\hline
\noalign{\smallskip}
 675.6851   &   8 & $^5\mathrm{P}_3-{^5}\mathrm{D}_2^\mathrm{o}$ & --1.76  & 7.87    & --7.146\\
 675.7007   &   8 & $^5\mathrm{P}_3-{^5}\mathrm{D}_3^\mathrm{o}$ & --0.90  & 7.87             & --7.146\\
 675.7171   &   8 & $^5\mathrm{P}_3-{^5}\mathrm{D}_4^\mathrm{o}$ & --0.31  & 7.87             & --7.146\\
 869.3931   &   6 & $^5\mathrm{P}_3-{^5}\mathrm{D}_3^\mathrm{o}$ & --0.51  & 7.87             & --7.337\\
 869.4626   &   6 & $^5\mathrm{P}_3-{^5}\mathrm{D}_4^\mathrm{o}$ &   0.08  & 7.87             & --7.337\\
1045.5449 &   3 & $^3\mathrm{S}_1^\mathrm{o}-{^3}\mathrm{P}_2$ &   0.26  & 6.86             & --7.672\\
1045.6757 &   3 & $^3\mathrm{S}_1^\mathrm{o}-{^3}\mathrm{P}_0$ & --0.43  & 6.86             & --7.672\\
1045.9406 &   3 & $^3\mathrm{S}_1^\mathrm{o}-{^3}\mathrm{P}_1$ &   0.04  & 6.86             & --7.672\\
\noalign{\smallskip}
\hline
\end{tabular}
\end{center}
\end{table}


\section{Models}

For the Sun and Procyon our analysis is based on 3D hydrodynamical model
atmospheres computed with the \cobold\ code \citep{2002AN....323..213F,2004AA...414.1121W}.
\cobold\ solves the coupled non-linear equations of compressible hydrodynamics
in an external gravity field together with non-local frequency-dependent
radiation transport for a small volume located at the stellar surface (see
\cobold\ manual
\href{http://www.astro.uu.se/~bf/cobold/index.html}{http://www.astro.uu.se/$\sim$bf/cobold/index.html}).
25 snapshots were selected from a \cobold\ simulation to represent the
photosphere of the Sun (Caffau et al., accepted for A\&A Letters),
with an effective temperature of 5780\pun{K} and covering 6000\pun{s} of
temporal evolution. 28 snapshots were selected from a 3D simulation of
Procyon, having an effective temperature of 6500\pun{K} and covering a time
interval of 16800\pun{s}. Since the timescale of the evolution of the
granular flow is about 2.8~times longer in Procyon than in the Sun, the 
simulated time spans are very similar in a dynamical sense.
The Procyon model used in this paper 
is the same as
the one used by \citet{2005ApJ...633..424A} in their study of the star's
centre-to-limb variation.
In both hydrodynamical models, an opacity binning scheme with five
wavelength bins was applied for modelling the wavelength-dependence of the
radiative transfer \citep{1982A&A...107....1N,1994A&A...284..105L,2004A&A...421..741V}.

For the other stars we do not have available 3D atmospheres, and therefore
used 1D LTE plane-parallel models.  Such 1D models were also used as a 
reference for the Sun and Procyon. In particular:

\begin{enumerate}

\item For all the stars we computed hydrostatic 1D model atmospheres computed
  with the LHD code. LHD is a Lagrangian 1D (assuming plane-parallel geometry)
  hydrodynamical model atmosphere code. It employs the same micro-physics
  (equation-of-state, opacities) as \cobold. The convective energy transport
  is described by mixing-length theory. The spatial discretisation and
  numerical solution of the radiative transfer equation is similar to the one
  in \cobold, albeit simplified for the 1D geometry. The
  wavelength-dependence of the radiation field is treated by the opacity
  binning method. A hydrostatic stratification in radiative-convective
  equilibrium is obtained by following the actual thermal and dynamical 
  evolution in time of the atmosphere until a stationary state is reached.  LHD
  produces standard 1D model atmospheres which are differentially comparable
  to corresponding 3D \cobold\ models.  Remaining choices entering an LHD
  model calculation are the value of the mixing-length parameter, which
  formulation of mixing-length theory to use, and in which way turbulent
  pressure is treated in the momentum equation.  Note, that these degrees of
  freedom are also present in other 1D model atmosphere codes.  The LHD models
  presented in this work have all been computed with $\alpha_{\rm MLT}=1.5$
  using the formulation of \citet{1978stat.book.....M}, and turbulent pressure
  has been neglected. Comparisons to 3D models were always made with
  LHD models having the same effective temperature, gravity, and chemical
  composition as the 3D model.

\item For all stars, except the Sun, we
 computed  ATLAS9 
 \citep{1993KurCD..13.....K,2005MSAIS...8...14K}
models
using 
the Linux version \citep{2004MSAIS...5...93S,2005MSAIS...8...61S} of the code.
 All these models have been computed with the
``NEW'' Opacity Distribution Functions 
\citep{2003IAUS..210P.A20C}, 
which are based on solar abundances from \citet{1998SSRv...85..161G}
with 1\kms\ micro-turbulence,  a
mixing-length parameter~\mlp\ of 1.25
and no overshooting.

\item 
We used
an   ATLAS9 
\citep{1993KurCD..13.....K,2005MSAIS...8...14K} model of the Sun computed
  by Fiorella Castelli with the solar abundances of
  \citet{2005ASPC..336...25A} as input for the chemical composition.  The
  Opacity Distribution Functions were explicitly computed for the same chemical
  composition and a micro-turbulent velocity of 1\kms.
  The model was computed assuming a mixing-length parameter of 1.25 and no overshooting.
  It is available at
  \href{http://wwwuser.oats.inaf.it/castelli/sun/ap00t5777g44377k1asp.dat}{http://wwwuser.oats.inaf.it/castelli/sun/
    ap00t5777g44377k1asp.dat}.

\item We used 
the Holweger-M\"uller solar model \citep{1967ZA.....65..365H,
    1974SoPh...39...19H}.  It is an empirical model of the solar photosphere
  and lower chromosphere assuming LTE, and largely reproducing the Sun's continuous and
  line spectrum.  It considers 900 selected line profiles of 31 atoms and
  ions.

\item For the Sun and Procyon we considered horizontal and temporal averages of
  the 3D snapshots over surfaces of equal (Rosseland) optical depth. 
  Comparison with these averaged 3D models, henceforth denoted as \mD\
  models, provide estimates of the influence of fluctuations around the mean
  stratification on the line formation process. Comparing 3D and 1D models of
  this kind is largely independent of arbitrary assumptions entering the 
  constructions of standard 1D models. The only
  free parameter which has to be specified for the 1D average model is the
  micro-turbulence to be used in related spectrum synthesis calculations.

\end{enumerate}

The spectral synthesis calculations for the \cobold, LHD, ATLAS and
Holweger-M{\"u}ller models were performed with the code \linfor\ (see
\href{http://www.aip.de/~mst/Linfor3D/linfor_3D_manual.pdf}{http://www.aip.de/$\sim$mst/Linfor3D/linfor\_3D\_manual.pdf});
for ATLAS models we also used  SYNTHE \citep{1993KurCD..18.....K,2005MSAIS...8...14K},
in its Linux version \citep{2004MSAIS...5...93S,2005MSAIS...8...61S}.
This was used in the cases in which
we wanted to compute  a synthetic spectrum containing many lines
from other atoms and molecules, to fit the observed data. 
The present version of  
\linfor\ can handle at maximum a few  tens of lines at a time,
while SYNTHE does not suffer such a limitation.


\section{3D abundance corrections}

The purpose of analysing a star both with 3D hydrodynamical
models and 1D models is to derive ``3D abundance corrections''
which may be used to correct the analysis
of other stars of similar atmospheric parameters, performed
in 1D. It is clear that from the computational point
of view 1D modelling (both model atmosphere and spectrum
synthesis) is much more convenient than 3D modelling.
It is unlikely that in the near future the computing
power will increase to the level that 3D analysis of
stellar spectra will be done routinely. 
On the other hand it is conceivable that grids of
hydrodynamical models will be constructed from
which ``3D corrections'' can be computed, and these 
can be used to correct the results of a 1D analysis.  
It is therefore important to define what is meant
by a ``3D correction''.
In this paper we give 3D-1D abundance corrections with
reference to LHD models.  These provide a reasonably well-defined way to
establish the relation between a 3D hydrodynamical and 1D hydrostatic model.
The fact that LHD and \cobold\ use the same opacities and
micro-physics ensures that the differences reflect {\em only}
the 3D effects and no other effects, as would be the
case if we computed corrections with respect to other
hydrostatic models, like ATLAS.
It may be useful to 
summarise  the differences between a 3D hydrodynamic model
and a 1D hydrostatic model as due to two factors:
\begin{enumerate}
\item
a different mean structure
of the two classes of models;
\item
horizontal temperature and pressure fluctuations,
which are present in a 3D simulation
but not (by definition) in a 1D model.
\end{enumerate} 
Our definition of ``3D correction'' is made in order to take into account both
effects at the same time. 

If one wants to see the difference in abundance
due {\em only} to the different mean temperature structure
one may compare the abundances derived from the
horizontally and temporally averaged 3D model, which we defined above
as the \mD\ model, to the corresponding LHD model.
On the other hand, if one wants to evaluate {\em only} the 
effect of horizontal temperature and fluctuations,
it is more appropriate to compare the 3D abundances with
those derived from the \mD\ model.


\section{Data}

For the Sun
we considered two high resolution, high signal to noise
ratio, spectra of the solar flux.

\begin{enumerate}

\item
The spectrum we will refer to as ``Kurucz flux'' 
is based on fifty solar Fourier transform spectrometer (FTS)
scans taken by J.~Brault and L.~Testerman at Kitt Peak,
with a spectral resolution of the order of 300\,000 and signal 
to noise of around 3\,000,
varying from range to range (further
details can be found in  \citealt{2005MSAIS...8..189K}).

\item
The ``Neckel flux'' refers to
the \citet{1984SoPh...90..205N}
absolutely calibrated 
FTS spectra obtained at Kitt Peak, 
covering the range 330\pun{nm} to 1250\pun{nm}.  
The  spectral purity  ranges from 0.4\pun{pm} at  330\pun{nm}
to 2\pun{pm} at 1250\pun{nm}. This means that the resolution at
1045\pun{nm} is about 500\,000.
\end{enumerate}

The spectra for Procyon and the other stars were
obtained from the UVES Paranal Observatory Project \citep{2003Msngr.114...10B}.
For the 670\pun{nm} and the 870\pun{nm} lines, we have taken the reduced data
present in the UVES POP web site\footnote{
\href{http://www.sc.eso.org/santiago/uvespop/}{http://www.sc.eso.org/santiago/uvespop/}
}, the
spectral resolution is of about 80\,000. 
The lines of Mult.~3
are not available in the POP reduced spectra, although
inspection of the raw data reveals that these wavelengths
are indeed recorded on the MIT CCD of UVES
in the standard 860\pun{nm} setting, but
the number of counts is very low (UVES is very inefficient
at these wavelengths) and only  about 1/3 of the order is 
present on the CCD.
For these two reasons a standard extraction with
the UVES pipeline fails to extract this order.
We downloaded the raw data and associated calibration
observations from the ESO archive and reduced them
using the UVES pipeline.
We forced the extraction  of the last order in 
the 860\pun{nm} setting
by declaring the number of orders to be found.
This allowed a satisfactory order definition
and subsequent order extraction.


\section{Data analysis}
\label{s:dataanalysis}

To derive sulphur abundances
we measured the equivalent width (EW) 
of the selected sulphur lines
and, when possible, fitted the line profiles.

For the triplet lines, EWs were computed by direct numerical integration 
using the trapeze sum rule,
or with the integration of a fitted
Gaussian for weak lines or Voigt profile for strong lines, 
using the IRAF task {\tt splot}.

All line profile fitting has been performed using
a code, described in  \citet{2005A&A...441..533C}, which performs
a $\chi ^2$ minimisation of the deviation between synthetic profiles
and the observed spectrum.
In the fitting, the sulphur abundance, 
the level
of the continuum, and a wavelength shift are left as free parameters,
while the macro-turbulence has to be  fixed {\em a priori}.

When available, 
for the Van der Waals broadening we used the parameter
derived from the theory by
\citet{1995MNRAS.276..859A,1997MNRAS.290..102B,1998MNRAS.296.1057B} and summarised in 
\citet{1998PASA...15..336B}; in this paper we collectively refer to this work
as ``ABO'' theory.
Otherwise we relied on the  approximation 
build into the SYNTHE spectrum synthesis suite
\citep{1993KurCD..18.....K,2005MSAIS...8...14K}.
The  comparison of this approximation 
to other ones can be found in
\citet{1998AA...331.1051R}, who refers 
to its use in the WIDTH code \citep{2005MSAIS...8...14K}.
We note here that the approximation is used in SYNTHE
and WIDTH for all lines for which literature
data on the Van der Walls broadening do not exist, and not
only for iron-peak elements, as the reader might be
induced to believe reading the paper of \citet{1998AA...331.1051R}.

We used a 
3D \cobold\ simulation for the sulphur analysis in the 
Sun and Procyon. 
For $\epsilon$ Eri (HD~22049) we computed the 3D-1D abundance corrections 
by using a \cobold\ simulation whose parameters (\Teff=5073\pun{K}, 
\glog=4.42, \moh=0.0) were very close to its stellar parameters.


\subsection{Corrections for departure from local thermodynamic equilibrium}

According to \citet{2005PASJ...57..751T} several of the atomic sulphur lines show
non-negligible departures from LTE.
It is beyond the purpose of this paper to investigate
these departures, however we shall make use
of the published departures from LTE of 
\citet{2005PASJ...57..751T}. 


\section{Results for the individual stars}

\subsection{Sun}

For the Sun we consider the 675.7\pun{nm} triplet of Mult.~8,
the 869\pun{nm} lines of Mult.~6 and the three lines of Mult.~3 even if we know
that the bluest is blended.

The results are reported in Table \ref{sunzolfo} and are: the EW for each
feature (second column), the sulphur abundance derived from the EWs and the
results for the 3D corrections, assuming a micro-turbulence $\xi_{\rm micro}$
of 1.5\kms\ and 1.0\kms\ in the 1D models.  The 3D corrections are always defined
with respect to the corresponding 1D LHD model.  In LTE, the 3D abundance
corrections for the Sun turn out to be small. By magnitude this is compatible
with the findings of \citet{2004AA...415..993N}, however, our corrections have
opposite sign. It is quite likely that the difference comes about by the
different choice of 1D reference atmosphere to which the corrections are
related. 
In fact, \citet{2004AA...415..993N} used a MARCS model as the 1D 
reference. 
In the table the NLTE correction, according to
\citet{2005PASJ...57..751T}, is included in the last but one column
($\Delta$).  The last column is the adopted sulphur abundance.  In columns
4-11 there are the sulphur abundance values derived from 1D models, using a
$\xi_{\rm micro}$ of 1.0 and 1.5\kms.

\begin{figure*}
\resizebox{\hsize}{!}{\includegraphics[clip=true,angle=0]{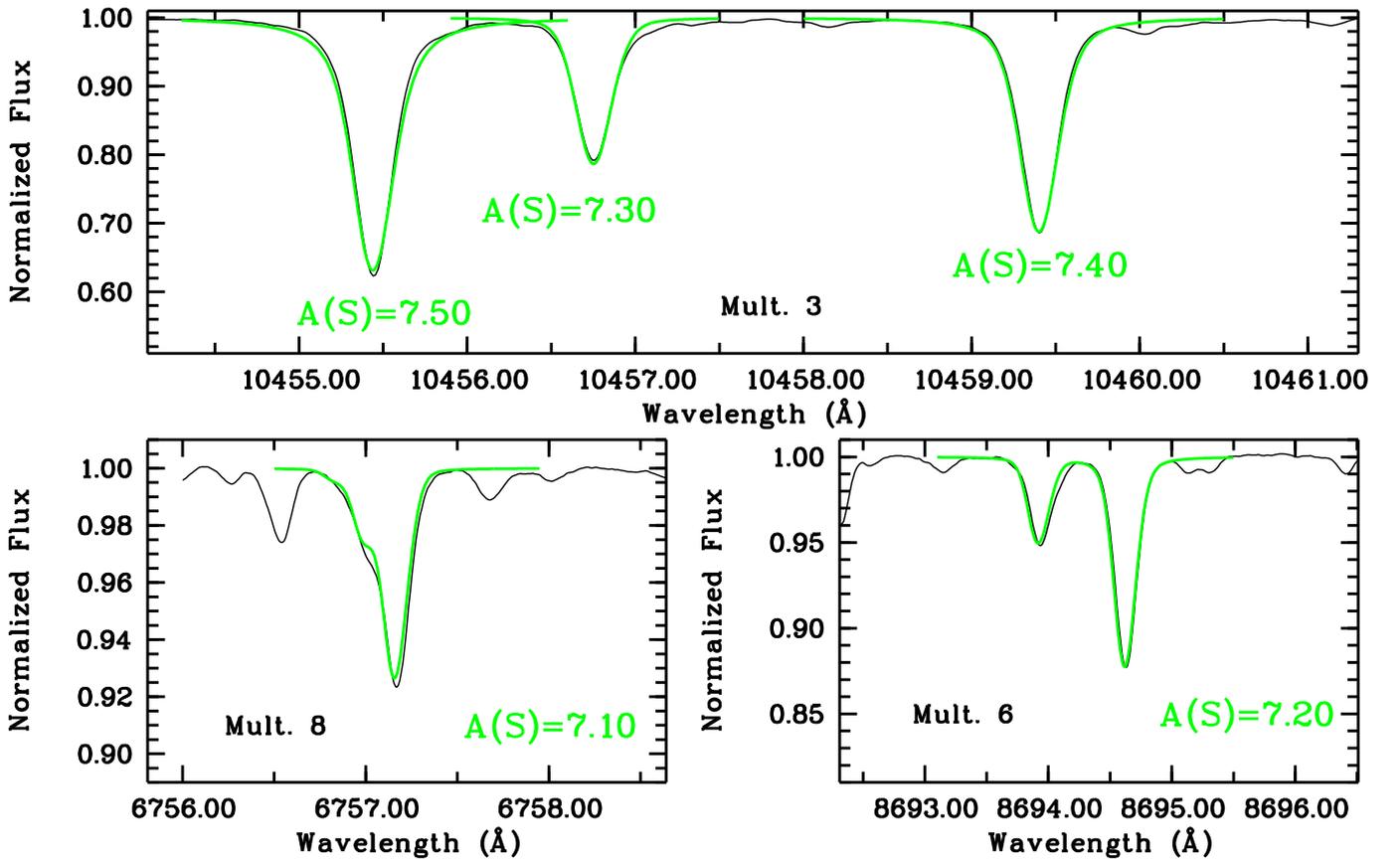}}
\caption{Sun: synthetic spectra (green/grey thick solid line) based on the 3D
  model atmosphere are over-imposed on the
observed (black solid line) solar flux spectrum of Kurucz.
}
\label{sunlines}
\end{figure*}

In Fig. \ref{sunlines} the solar sulphur lines considered in this
work are plotted together with the 3D synthetic spectra.
The weak lines of Mult.~6 and Mult.~8 have also been fitted with both the 
\cobold+\linfor\ synthetic spectra and ATLAS9+SYNTHE. 
The results are, within 0.05\,dex, in agreement
with the abundances obtained from the EWs.

\begin{table*}
\begin{center}
\caption{Solar sulphur abundances from flux spectra.
Col. (1) is the wavelength of the line followed
by an identification flag, K means Kurucz flux, N Neckel flux;
col. (2) is the Equivalent Width;
col. (3) is the sulphur abundance, A(S), according to the \cobold\ 3D model;
col.~(4) to (11) give the abundances for a micro-turbulence of 1.5 and
1.0\kms\ 
for the \mD, ATLAS, HM, and LHD model, respectively;
col. (12) and (13) provide the 3D abundance corrections relative to the LHD
model for a
micro-turbulence of 1.5 and 1.0\kms, respectively;
col. (14) is the NLTE correction from \citet{2005PASJ...57..751T};
col. (15) the A(S) we finally adopted.
LHD models were
computed with
\mlp\ of 1.5,  ATLAS  models  with \mlp\ of 1.25.
\label{sunzolfo}}
\begin{tabular}{rrrrrrrrrrrrrrr}
\hline
\noalign{\smallskip}
Wave      &  EW    & \multicolumn{9}{c}{A(S) from EW} & \multicolumn{2}{c}{3D-LHD} & $\Delta$& A(S)\\
nm        &  pm    & 3D   &  \multicolumn{2}{c}{\mD} & \multicolumn{2}{c}{ATLAS} & \multicolumn{2}{c}{HM} & \multicolumn{2}{c}{LHD} \\
          &        &      &   1.5  &   1.0  &   1.5  &   1.0  &   1.5  &   1.0  &   1.5  &   1.0 &  1.5  &   1.0 \\  
 (1)      &  (2)   & (3)  &  (4)   &  (5)   &  (6)   &  (7)   &  (8)   &  (9)   &  (10)  &  (11) & (12)&(13)& (14) & (15) \\
\noalign{\smallskip}
\hline
\noalign{\smallskip}
  675.7K&   1.805&   7.138&   7.142&   7.151&   7.130&   7.137&   7.168&   7.176&   7.121&   7.129&  0.017&   0.009&        &   7.138\\
  675.7N&   1.789&   7.133&   7.137&   7.146&   7.125&   7.132&   7.163&   7.171&   7.116&   7.124&  0.017&   0.009&        &   7.133\\
  869.3K&   0.844&   7.028&   7.036&   7.042&   7.028&   7.033&   7.059&   7.064&   7.018&   7.023&  0.010&   0.005&  -0.010&   7.018\\
  869.3N&   0.863&   7.040&   7.048&   7.054&   7.040&   7.045&   7.070&   7.076&   7.030&   7.035&  0.010&   0.005&  -0.010&   7.030\\
  869.4K&   3.025&   7.189&   7.177&   7.196&   7.153&   7.169&   7.182&   7.202&   7.144&   7.161&  0.045&   0.028&  -0.010&   7.179\\
  869.4N&   3.157&   7.219&   7.206&   7.225&   7.182&   7.198&   7.215&   7.236&   7.173&   7.190&  0.046&   0.029&  -0.010&   7.209\\
 1045.5K&  13.660&   7.402&   7.320&   7.378&   7.277&   7.330&   7.346&   7.405&   7.266&   7.321&  0.136&   0.081&  -0.090&   7.312\\
 1045.5N&  13.450&   7.383&   7.302&   7.360&   7.259&   7.312&   7.328&   7.386&   7.247&   7.303&  0.136&   0.080&  -0.090&   7.293\\
 1045.6K&   6.872&   7.387&   7.326&   7.374&   7.298&   7.342&   7.333&   7.382&   7.288&   7.332&  0.099&   0.055&  -0.050&   7.337\\
 1045.6N&   6.725&   7.366&   7.306&   7.353&   7.279&   7.322&   7.313&   7.362&   7.269&   7.312&  0.098&   0.054&  -0.050&   7.316\\
 1045.9K&  10.650&   7.370&   7.292&   7.347&   7.254&   7.306&   7.307&   7.364&   7.244&   7.298&  0.126&   0.073&  -0.070&   7.300\\
 1045.9N&  10.610&   7.366&   7.288&   7.343&   7.250&   7.302&   7.303&   7.360&   7.240&   7.293&  0.126&   0.073&  -0.070&   7.296\\
\noalign{\smallskip}
\hline
\end{tabular}
\end{center}
\end{table*}

The LTE computation implies A(S)=$7.252\pm 0.143$ (this 
is an average of all the A(S) given in 
column (3) of Table \ref{sunzolfo}), while applying 
the NLTE corrections
of \citet{2005PASJ...57..751T} (given in column (11) of
Table \ref{sunzolfo})  we obtain
A(S)=$7.213\pm 0.113$; 
the quoted errors 
are the line to line scatter in the
abundance determinations from the six features considered.
It is worth pointing out that, given the high 
S/N of the available solar spectra, the associated statistical
errors are negligible. Therefore the line to line scatter
must reflect inadequacies in the model atmospheres
and/or the line formation calculations,
and/or  errors in the atomic data.

Examining more closely  the  disagreement between lines, we note that
the 869.3\pun{nm} line is blended with molecular lines 
(C$_2$ and CN)
whose contribution we estimated to be  
of the order  of 15\,\%, and which was consequently   subtracted from the EW. 
However, considering the uncertainties in the \loggf\ of these
molecular lines, it is probably safer  to discard this line.
From the 675.7\pun{nm} triplet  we find A(S)=$7.136\pm 0.004$ 
(where the error is now the standard deviation between
the measures from the two observed solar spectra),
while the sulphur abundance is $7.204\pm 0.021$ in LTE
and $7.194\pm 0.021$ in NLTE for the 869.4\pun{nm} line. 
From Mult.~3 we find 
A(S)=$7.379\pm 0.014$ in LTE and A(S)=$7.309\pm 0.016$ in NLTE.
There is an obvious trend in A(S) with EW. Note that 
since this result is obtained using 3D atmospheres 
we cannot invoke a micro-turbulent velocity to remove this trend. 
This effect could be explained if
3D-NLTE corrections are larger than the
published  1D-NLTE corrections, adopted here.
If this were the case, the solar S abundance should be A(S)=7.14,
as indicated by the 675.7\pun{nm} triplet, which is virtually unaffected by
NLTE. This result is also supported by the analysis of [SI] line at 1082\pun{nm}
(Caffau \& Ludwig, in press),
whose departure from LTE is negligible, from which the sulphur abundance
is 7.14.
However, the trend could have other explanations. 
At least in part, the line broadening theory employed could be inadequate,
as could be the case for 
the hydrodynamic velocity and temperature field of the \cobold\ simulation.
Finally, we note that \citet{2005ASPC..336...25A} find a solar LTE
sulphur abundance of A(S)=$7.14 \pm 0.05$.


\subsection{Procyon}

For Procyon we adopt the stellar parameters derived by  
\citet{1985AAS...59..403S},
that is \teff = 6500\pun{K}, \glog=4.0 and solar metallicity.
These stellar parameters are very close to the
results of the most recent determination due to  \citet{2005ApJ...633..424A}, 
who find \Teff=$6516\pm 87$~K, \glog=$3.95\pm 0.02$.
For Mult.~8 and Mult.~6 we measure S/N=400, for Mult.~3 S/N=50.

For the 3D fitting we assumed a Gaussian instrumental broadening
with FWHM of 3.75\kms, corresponding to a spectral resolution of 80\,000.

\begin{figure}
\resizebox{\hsize}{!}{\includegraphics[clip=true,angle=0]{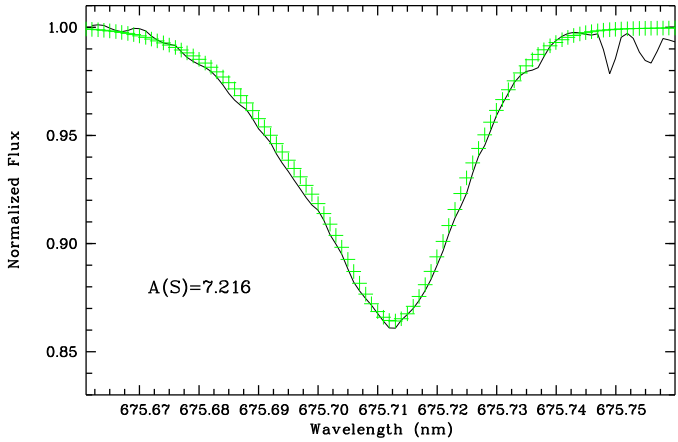}}
\caption{Procyon: the 3D fit (green/grey crosses) is over-imposed on the observed spectrum 
(black solid line) for the \ion{S}{I} 675.7\pun{nm} triplet.
}
\label{prociofit}
\end{figure}

The results are reported in Table \ref{procio}.
In the second column the EW is given, the next five
columns show the sulphur abundance computed from EWs.
The next two columns show the sulphur abundance from
line profile fitting (see in Fig. \ref{prociofit} the fit
of the 675.7\pun{nm} triplet); the two lines (869.3931\pun{nm} and
869.4626\pun{nm} line) of Mult.~6 are
fitted at the same time.
In column (10) the 3D-1D correction is given,
in column (11), when available, the NLTE correction
according to \citet{2005PASJ...57..751T} is reported. 
In the last but one column the adopted sulphur abundance is given.

The standard deviation of the EW related to the finite S/N is computed according to Cayrel's formula
\citep{1988IAUS..132..345C}: 
$$\epsilon\left(\mathrm{EW}\right)=1.6\times{\sqrt{{\rm FWHM} \times {\rm PixelSize}}\over S/N}\eqno{(1)}$$
where $S/N$ is the signal to noise value, FWHM is the full width at half maximum of the line,
PixelSize is the size of the detector pixel 
in wavelength  units.
From the errors of the EWs we computed 
the corresponding errors of A(S) using the curve of growth
of each line. These errors  are 
provided in the last column of Table \ref{procio}.

\begin{table*}
\begin{center}
\caption{Sulphur abundances in Procyon.
Col. (1) is the wavelength of the line;
col. (2) is the Equivalent Width;
col. (3) is the sulphur abundance, A(S), according to the \cobold~3D model;
col. (4) is A(S) from the \mD\ model;
col. (5) is A(S) from the ATLAS 1D model;
col. (6) is A(S) from the LHD 1D model;
col. (7) and (8) are the A(S) from fitting using a \cobold~3D
and a ATLAS+SYNTHE grid, respectively;
in all 1D models a micro-turbulence of 2.1\kms\ was assumed;
col. (9) is the 3D correction;
col. (10) is the NLTE correction from \citet{2005PASJ...57..751T};
col. (11) is the A(S) we adopted;
col. (12) is the statistical error.
\mlp\ is of 1.50 for the LHD   model and 1.25 for the 
ATLAS model.
\label{procio}}
\begin{tabular}{rrrrrrrrrrrr}
\hline
\noalign{\smallskip}
Wavelength&  EW    & \multicolumn{4}{c}{A(S) from EW} & \multicolumn{2}{c}{A(S) from fit}& 3D-LHD & $\Delta$ & A(S)& $\sigma$\\
 (nm)     &  (pm)  &  3D   & \mD    &  ATLAS & LHD & 3D & ATSY \\
 (1) & (2) & (3) & (4) & (5) & (6) & (7) & (8) & (9) & (10) & (11) & (12) \\
\noalign{\smallskip}
\hline
\noalign{\smallskip}
\\
 675.7 &  4.371 &  7.231&  7.197&  7.141&  7.177&  7.216&  7.162&  0.053&       &  7.231 & 0.005\\
 869.3 &  3.182 &  7.332&  7.298&  7.239&  7.273&  7.257&  7.214&  0.059& -0.038&  7.294 & 0.005\\
 869.4 &  6.007 &  7.265&  7.187&  7.120&  7.138&       &       &  0.128& -0.057&  7.268 & 0.005\\
1045.5 & 18.000 &  7.557&  7.352&  7.326&  7.274&       &       &  0.284& -0.271&  7.286 & 0.030\\
1045.6 & 10.180 &  7.410&  7.242&  7.193&  7.178&       &       &  0.233& -0.135&  7.275 & 0.040\\
1045.9 & 15.370 &  7.547&  7.334&  7.302&  7.258&  7.199&       &  0.289& -0.231&  7.316 & 0.030\\
\noalign{\smallskip}
\hline
\noalign{\smallskip}
\end{tabular}
\end{center}
\end{table*}

For Procyon the 3D corrections are definitely not negligible, above all for
the strong lines of Mult.~3. This effect is not so evident in the Sun which is
cooler than Procyon. We shall see that the 3D corrections are negligible in
the cooler $\epsilon$ Eri (see Table \ref{hd22049}).  
We note that the 3D corrections given in Tab.~\ref{procio} are positive,
implying that -- for a given abundance -- the same lines are weaker in 3D 
than in 1D. This behaviour is opposite to the findings by
\citet{2002AA...387..258S} who argue that in general one should expect a
strengthening of lines in 3D atmospheres. However, their argumentation
is not strict, and actually refers to the 3D-\mD\ correction for
weak (unsaturated) lines in the Sun. In fact, the weaker lines in 
Tab.~\ref{sunzolfo} comply with the expected line strengthening in 3D.

The fact that the difference of the 3D abundance to that derived from 
the \mD\ model (col.~(3) -- col.~(4) in Tab.~\ref{procio}) is of 
similar size as the total 3D-1D correction given in col.~(9) 
(col.~(3) -- col.~(6) in Tab.~\ref{procio}) shows that the 
pronounced line weakening cannot be the result of a different mean
temperature structure in 3D with respect to 1D. In fact the abundances 
derived from the \mD\ (col.~(4)) and the LHD model (col.~(6)) are 
very similar. Rather the line weakening must be caused by the sizable
horizontal temperature fluctuations present in the photosphere of Procyon,
or by a substantially different effective micro-turbulence in 3D and 1D,
respectively.

\begin{figure}
\resizebox{\hsize}{!}{\includegraphics[clip=true,angle=0]{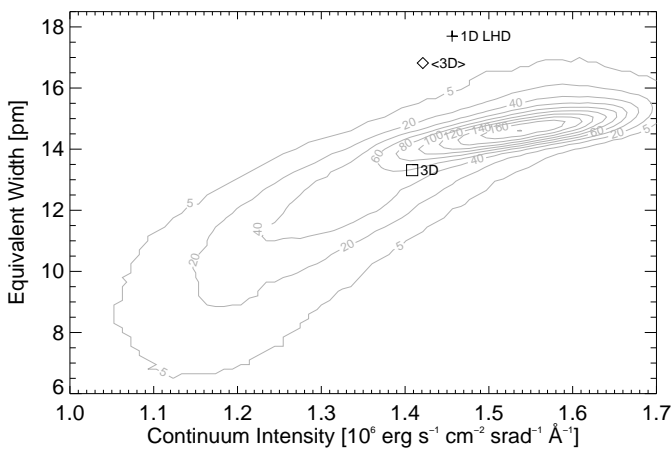}}
\caption{Procyon: joint probability density function (not normalised) of continuum
  intensity and (disk-centre) EW for the 1045.9\pun{nm} line 
in the 3D model (grey contours). The labelled
symbols mark the 3D and \mD\ average, as well as the result for a 
LHD model with \mlp=1.5. The micro-turbulent velocity for the 1D 
models is 2.1\kms.}
\label{ewprocio}
\end{figure}

\begin{figure}
\resizebox{\hsize}{!}{\includegraphics[clip=true,angle=0]{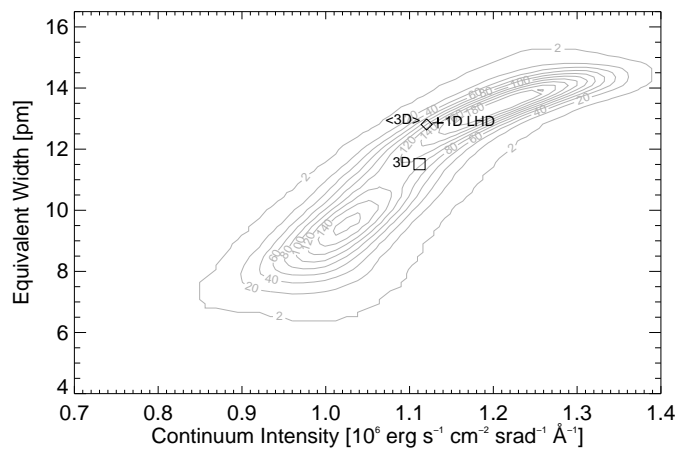}}
\caption{Sun: same as Fig.~\ref{ewprocio} for the solar case.
The micro-turbulent velocity  for the 1D models is 1.0\kms.}
\label{ewsun}
\end{figure}

In order to explore this question somewhat further, Figures~\ref{ewprocio} 
and~\ref{ewsun} depict the joint probability density of the (disk-centre) 
EW of the 1045.9\pun{nm} line and the neighbouring continuum intensity for
the Procyon and solar 3D model, respectively. The contours illustrate the
correlation between continuum intensity and EW over the stellar surface. We
find a bimodal distribution in the solar case, a single peak and diffuse
``halo'' in the case of Procyon. The difference can be traced back to
qualitatively different formation heights of the line. In the case of the Sun
the largest contribution to the line absorption stems from the granular layers
as such, while the maximum contribution is shifted to higher layers of reverse
granulation 
in Procyon. While in the Sun temperature fluctuations in the continuum
formation layers are positively correlated with the temperature fluctuations
in the line forming layers, we typically find an anti-correlation in Procyon.
Due to the high excitation potential of the lower level the line is
rather temperature sensitive. In the Sun instances of high continuum intensity
coincide with a high temperature in the line forming layers. Higher
temperatures increase the population of the lower level of the transition
leading to a stronger line. The reverse happens in moments of low continuum
intensity. In Procyon, the line still tends to become stronger with
increasing continuum intensity. However, the anti-correlation between
continuum intensity and temperature in parts of the line forming layers
leads to a smaller variation of the line strength.

The \mD\ average and the LHD model indicated by symbols in 
Fig.~\ref{ewprocio} point 
towards a potential problem in our determination of 3D abundance corrections 
for Procyon. None of the about $6\times 10^4$ vertical profiles entering the
calculation of the probability density corresponds to the \mD\ model
-- at least as far as it concerns the line formation properties. The same
holds for the LHD model. It appears plausible that the micro-turbulence
prescribed in 1D line formation calculations -- motivated from observations --
is too large in comparison to the effective micro-turbulence intrinsic to the
3D model. 
We note here that our adopted micro-turbulence of 2.1\kms\ from 
\citet{1985AAS...59..403S} is in agreement, within quoted errors,
with those of \citet{1998AA...338..161F,1996AA...314..191G,1998PASJ...50..509T}
and \citet{2002ApJ...567..544A}, while \citet{1996PASJ...48..511T} prefer a
lower value of 1.4\kms.
Test calculations have shown that a reduction on
$\xi_\mathrm{micro}$ from 2.1\kms\ to 1.5\kms\ leads to a reduction of the
abundance correction 3D-LHD by a factor of two. In part, this may explain the
different 3D corrections we obtain in comparison to \citet{2004AA...415..993N}
who find negligible corrections for an only slightly cooler atmosphere
(\Teff=6191\pun{K}, \logg=4.04, \moh=0.0). On the other hand, all sulphur lines
which span a range of a factor five in equivalent width provide a consistent
abundance in 3D when 1D-NLTE effects are included. We cannot provide a
resolution here, but the issue of the appropriate micro-turbulence in 3D-1D
comparisons clearly needs further investigation.

In LTE we obtain a sulphur abundance of $7.400\pm 0.131$, while including NLTE
corrections it becomes $7.278\pm 0.029$.  Once NLTE corrections are applied
the scatter becomes tiny and fully compatible with the noise in the data.  We
commented before that the 869.3\pun{nm} line is blended by molecular lines.
If we discard this line A(S)=$7.250\pm 0.026$.  Thus, the difference is small.
This is perhaps not surprising, since Procyon is hotter than the Sun, and
molecular lines are less prominent in this spectral range.

\begin{figure*}
\resizebox{\hsize}{!}{\includegraphics[clip=true,angle=0]{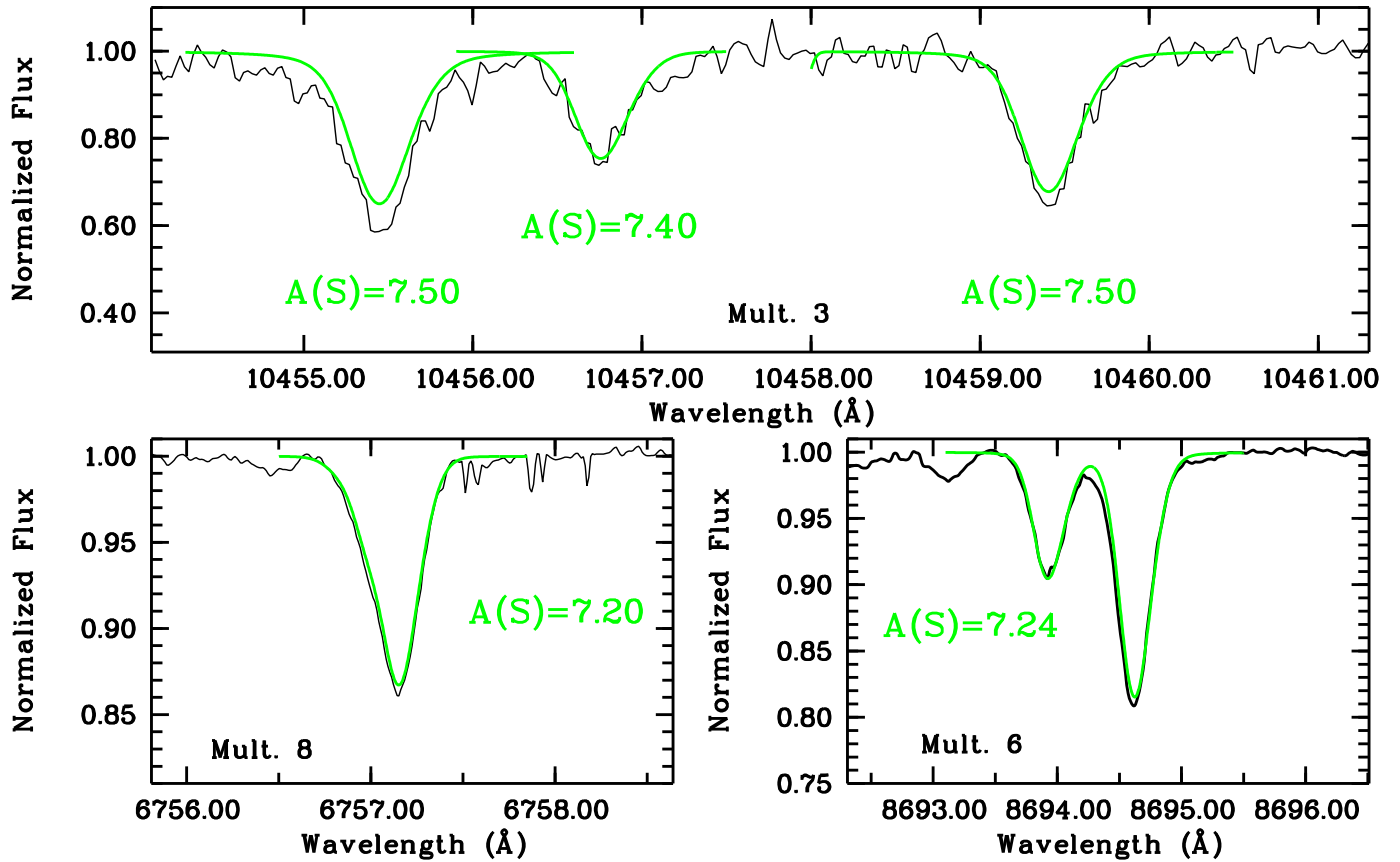}}
\caption{Procyon: synthetic spectra (green thick line) based on the 3D model
  are over-imposed on the observed (black solid line) ones.
}
\label{sunline}
\end{figure*}


\subsection{HD 33256}

The stellar parameters found in the literature 
for this star show little scatter.
We assume \teff = 6\,454\pun{K}, which we derived
by fitting the  H$\alpha$ wings with a grid 
of synthetic spectra computed with ATLAS+SYNTHE.
We recall that the SYNTHE code computes the van der Waals broadening
of the Balmer lines according to the theory of 
\citet{1965PhRv..140.1044A}, and the Stark broadening
according to \citet{1973ApJS...25...37V}.
Had we adopted the broadening theory of
\citet{2000A&A...363.1091B} we would have obtained a lower
effective temperature, as pointed out e.g. by 
\citet{2007A&A...462..851B}.
We assumed the stellar parameters
log g = 4.00, \vsini = 10\kms, $\xi_{\rm micro}$ = 1.55\kms.
From  20 \ion{Fe}{i} lines we obtain [Fe/H] = $7.25\pm 0.11$;
18 lines of \ion{Fe}{ii} give [Fe/H] = $7.29\pm 0.15$.
For the ranges at 600\pun{nm} and 800\pun{nm} we measure  a S/N ratio
of about 700, and S/N = 25 around 1045\pun{nm}.

The results of our analysis are listed in Table \ref{hd33256}.
The errors of A(S), derived from these S/N ratios
according to the formula (1), 
are given in the last column of the table.

\begin{table*}
\begin{center}
\caption{Sulphur abundances in HD~33256.
Col. (1) is the wavelength of the line;
col. (2) is the Equivalent Width;
col. (3) is A(S) from ATLAS+\linfor;
col. (4) is A(S) from LHD+\linfor;
col. (5) is A(S) from fitting using an ATLAS+SYNTHE grid;
col. (6) is the NLTE correction from \citet{2005PASJ...57..751T};
col. (7) is the adopted A(S);
col. (8) is the statistical error of A(S).
\label{hd33256}}
\begin{tabular}{rrrrrrrr}
\hline
\noalign{\smallskip}
Wave      &  EW    & \multicolumn{3}{c}{A(S)}& $\Delta$ & A(S) & $\sigma
$\\
 (nm)     &  (pm)  &  & &  & FIT \\
          &        &  ATLAS & LHD\\
(1)&(2)&(3)&(4)&(5)&(6)&(7)&(8)\\
\noalign{\smallskip}
\hline
\noalign{\smallskip}
  675.7 &  2.37 & 6.822 & 6.863 & 6.802 &         & 6.863 & 0.004\\
  869.3 &  1.63 & 6.884 & 6.925 & 6.895 & --0.031 & 6.894 & 0.004\\
  869.4 &  3.92 & 6.844 & 6.872 & 6.895 & --0.044 & 6.828 & 0.004\\
 1045.5 & 16.74 & 7.349 & 7.300 & 7.219 & --0.239 & 7.061 & 0.079\\
 1045.6 &  7.65 & 6.977 & 6.974 & 7.022 & --0.159 & 6.815 & 0.116\\
 1045.9 & 13.35 & 7.236 & 7.196 & 7.119 & --0.258 & 6.938 & 0.081\\
\noalign{\smallskip}
\hline
\end{tabular}
\end{center}
\end{table*}

The LTE sulphur abundance is $7.022\pm 0.183$; applying
the \citet{2005PASJ...57..751T} 
NLTE corrections A(S)=$6.900\pm 0.091$.
After applying the NLTE corrections the line to line
scatter is fully compatible with
the expected measurement errors.
The sulphur abundance with NLTE correction from the
lines of Mult.~3 is $6.938\pm 0.123$,
to be compared with 
A(S)=$6.862\pm 0.033$ if we consider the 675.7\pun{nm},
869.3\pun{nm} and 869.4\pun{nm} lines, or
A(S)=$6.861\pm 0.047$ if we consider only the two lines of Mult.~6.
As can be seen, the sulphur abundances obtained from the
different lines are in agreement within one standard deviation.


\subsection{HD 25069}

We adopted the stellar parameter from
\citet{2005ApJS..159..141V}
(\teff = 4\,994\pun{K}, \glog ~= 3.53, and [Fe/H]=+0.10).

The fit of the 675.7\pun{nm} triplet gives A(S)=7.249, but, due to the
presence of a distortion in the line profile, the fit was performed in a range
not including the blue wing of the line.

S/N is 500 for the 600\pun{nm} and 800\pun{nm} ranges.
In the range 1045\pun{nm} we measure S/N=28.
EWs and sulphur abundances are reported in Table \ref{hd25069}.
The errors according formula (1) 
are provided  in the last column of Table \ref{hd25069}.

For this star the IR line at 1045.9\pun{nm}  appears 
to have an unusual shape,  the core appears broad and flat,
unlike the other atomic lines in this spectrum, for this 
reason we rejected this line.
The triplet at 675.7\pun{nm} should also be  rejected because 
the line profile appears distorted, possibly by a cosmic ray hit. 

\begin{table*}
\begin{center}
\caption{Sulphur abundances in HD~25069.
The format is identical to Tab.~\ref{hd33256}.}
\label{hd25069}
\begin{tabular}{rrrrrrrr}
\hline
\noalign{\smallskip}
Wave&  EW    & \multicolumn{3}{c}{A(S)} &$\Delta$ &A(S)& $\sigma$\\
 (nm) &  (pm)  & \multicolumn{2}{c}{linfor} & FIT \\
      &        & ATLAS & LHD\\
(1)&(2)&(3)&(4)&(5)&(6)&(7)&(8)\\
\noalign{\smallskip}
\hline
\noalign{\smallskip}
  675.7 &  1.35 & 7.277 & 7.290 & 7.249 &          & 7.290 & 0.010\\
  869.4 &  1.60 & 7.204 & 7.222 & 7.204 & --0.023  & 7.199 & 0.009\\
 1045.6 &  4.20 & 7.351 & 7.372 &       & --0.074  & 7.298 & 0.087\\
 1045.9 &  5.36 & 7.111 & 7.132 &       & --0.092  & 7.040 & 0.091\\
\noalign{\smallskip}
\hline
\end{tabular}
\end{center}
\end{table*}

For this star the contribution of the iron line to the 1045.5\pun{nm}
line of Mult.~3 is large.
So considering only the 869.4\pun{nm} and the 1045.6\pun{nm} line,
the sulphur abundance is $7.297\pm 0.106$, and
$7.248\pm 0.070$ if we apply the NLTE corrections of
\citet{2005PASJ...57..751T}.
Considering all lines A(S)=$7.254\pm 0.102$ in LTE,
A(S)=$7.207\pm 0.120$ with NLTE corrections of
\citet{2005PASJ...57..751T}.

Since  A(S)=7.199 from the 869.4\pun{nm} line,
and A(S)=$7.169\pm 0.182$ from the  the two lines considered in Mult.3, 
we can conclude that
the abundance determinations from the two lines are in good agreement.
We attribute the abundance difference of 0.26~dex between the lines
at 1045.6 and 1045.9~nm to the low S/N of the spectrum of Mult.~3.


\subsection{$\epsilon$ Eri (HD 22049)}

For this star the stellar parameters found
in the literature are again in good agreement.
We take the parameters of \citet{2004AA...415.1153S} 
(\teff = 5\,073\pun{K}, \glog ~= 4.42, $\xi_{\rm micro}$=1.05\kms and [Fe/H]=--0.13),
because these are also the parameters used by \citet{2004AA...426..619E} for
their sulphur abundance determination. The projected rotational velocity 
\vsini =3.0\kms\ stems from \citet{2004AA...418..989N}.

For this star we measure S/N=300 at 670\pun{nm} and a bit smaller than 300
at 870\pun{nm}, but
S/N=50 at 1045\pun{nm}.
The results are reported in Table \ref{hd22049}.
The 3D abundance correction given in column~(5) is the one for a star 
with \teff = 5000\pun{K}, \glog=4.44 and solar metallicity.

Also  for this star the contribution of the iron line to the 1045.5\pun{nm}
blend of Mult.~3 is high (12\,\% of the whole line),
therefore we exclude this line from the
S abundance determination.
The error of A(S) according to formula (1)
can be found in Table \ref{hd22049},
last column.

\begin{table*}
\begin{center}
\caption{Sulphur abundances in $\epsilon$ Eri (HD~22049).
Col. (1) is the wavelength of the line;
col. (2) is the Equivalent Width;
col. (3) is A(S) from ATLAS+\linfor;
col. (4) is A(S) from LHD+\linfor;
col. (5) is A(S) from fitting using an ATLAS+SYNTHE grid;
col. (6) is the 3D correction as 3D-LHD for a model of
         \teff = 5000\pun{K}, \glog~= 4.44 and solar metallicity;
col. (7) is the NLTE correction from \citet{2005PASJ...57..751T};
col. (8) is the adopted A(S);
col. (9) is the statistical error of A(S).
\label{hd22049}}
\begin{tabular}{rrrrrrrrr}
\hline
\noalign{\smallskip}
Wave&  EW    & \multicolumn{3}{c}{A(S)} & 3D-1D & $\Delta$ & A(S) & $\sigma$\\
 (nm)     &  (pm)  &  & & & FIT \\
          &        &  ATLAS & LHD\\
(1)&(2)&(3)&(4)&(5)&(6)&(7)&(8)&(9)\\
\noalign{\smallskip}
\hline
\noalign{\smallskip}
  675.7 &  0.62 & 7.162 & 7.234 & 7.080 & --0.0360 &         & 7.234 & 0.025\\
  869.4 &  1.02 & 7.183 & 7.252 & 7.047 & --0.0304 & --0.004 & 7.248 & 0.023\\
 1045.5 &  5.81 & 7.092 & 7.143 &       &   0.0268 & --0.047 & 7.096 & 0.035\\
 1045.6 &  2.52 & 7.175 & 7.240 &       & --0.0056 & --0.025 & 7.215 & 0.082\\
 1045.9 &  4.36 & 7.111 & 7.169 & 7.047 &   0.0033 & --0.037 & 7.132 & 0.057\\
\noalign{\smallskip}
\hline
\end{tabular}
\end{center}
\end{table*}

The LTE sulphur abundance is:
A(S)=$7.208\pm 0.048$, with NLTE correction $7.185\pm 0.067$.
A(S)=$7.148\pm 0.061$ if we consider the two lines of Mult.~3,
to be compared with  $ 7.241\pm 0.010$ from the  675.7\pun{nm}
and 869.4\pun{nm} lines.
For this star the agreement of the S abundance 
derived from the lines of Mult.~3 with that derived from other lines
is worse than for other stars.
We note also that  
the lines of Mult.~3
in the observed spectrum appear somewhat  distorted
and a higher S/N spectrum would be desirable to verify this result.


\section{Discussion}

\begin{figure}
\resizebox{\hsize}{!}{\includegraphics[clip=true,angle=0]{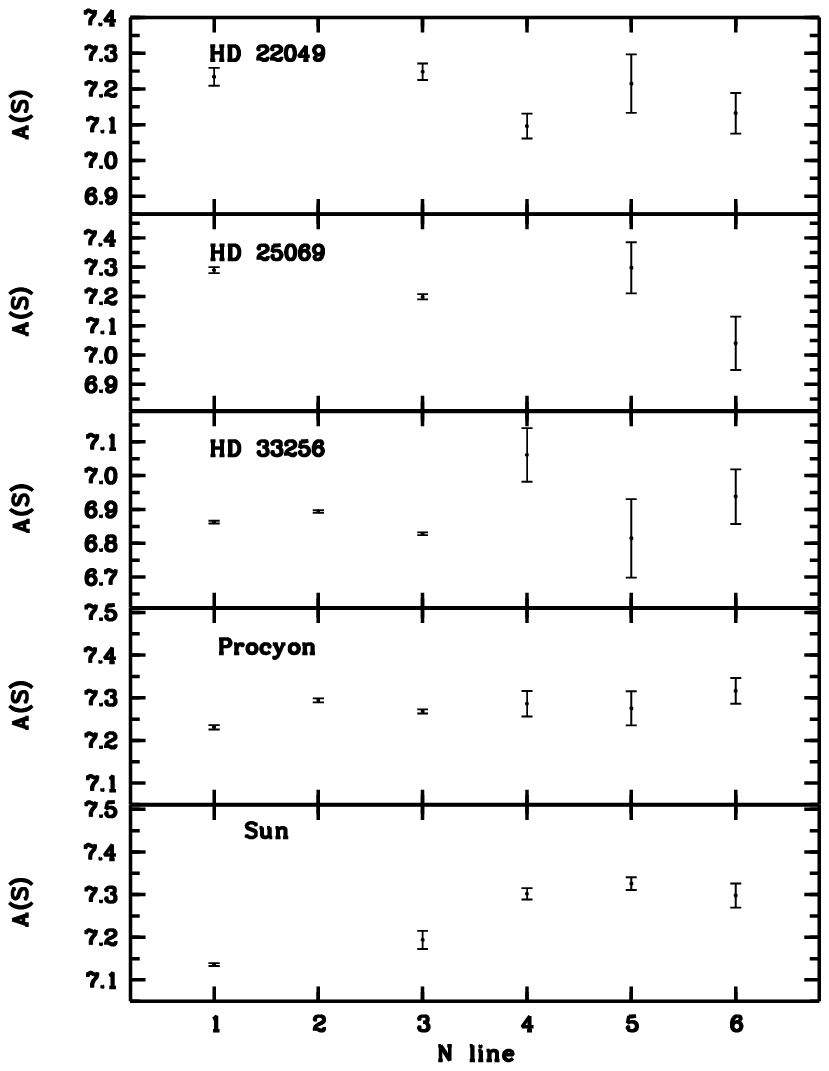}}
\caption{For each star, A(S) is plotted as a function of the number of
the line: 1 is the 675.7\pun{nm} triplet, 2 the 869.3\pun{nm} line,
3 the 869.4\pun{nm} line, 4 the 1045.5\pun{nm} line,
5 the 1045.6\pun{nm} line and 6 the 1045.9\pun{nm} line.
For the Sun the error bar is related to the statistical error
coming from the two observed flux spectra.
For the other stars it is related to the S/N through
Eq.~(1).}
\label{allstars}
\end{figure}

The main interest of this investigation
is the level of concordance  of the 
abundances derived from the lines of Mult.~3 with the
abundances derived from lines of other multiplets.
Our results are summarised in
 Fig. \ref{allstars}, where the value of A(S) for each
line is plotted for all stars.
There is no evident trend of A(S)
with respect to the line used to obtain the abundance,
except for the Sun, for which we have noted the
strong A(S)-EW correlation.
At this stage it is not clear if accounting for NLTE effects in the 3D model
may solve this problem.
However, the good agreement obtained for the other
stars is encouraging and suggests that \ion{S}{i} Mult.~3
is indeed a valuable abundance indicator.
The fact that these abundances are consistent 
for stars of different spectral types suggests
that all the oscillator strength ~ are on the same scale
and there is no systematic difference between 
the different multiplets. We repeat
that 
the error of the \loggf ~ for all lines used here is of
the order of
50\,\%, 
and better laboratory or theoretical oscillator
strengths would be highly desirable.

There is a tendency to underestimate the EWs when the observed spectrum is
broadened by an instrumental profile.
Tests on strong solar lines,
for which we have  a very high S/N spectrum
show a lower value for the measurements of the EW of more than 3\,\%,
when broadened to a resolution of R$\sim 80\,000$.
When dealing with
an  observed spectrum  of lower quality we expect this  effect to be
even larger.
According to our simulations an
IR low quality (S/N=20) spectrum can lead
to measure an EW of a strong line which is 10\,\% smaller than the true value,
because the wings are lost in the noise.

Our main conclusion is that Mult.~3 can be used successfully to
measure the sulphur abundance in the sample of stars we have considered in 
this work. We believe that this analysis can be extended to metal-poor 
stars, where the measurement of Mult.~3 lines with the IR spectrograph
CRIRES seems particularly promising.\\


\begin{acknowledgements}

PB and HGL acknowledge financial support from EU
contract MEXT-CT-2004-014265 (CIFIST). 
This research has made use of the SIMBAD database,
operated at CDS, Strasbourg, France.
Part of the hydrodynamical models used in this research
have been computed at CINECA, thanks to the INAF-CINECA
convention.

\end{acknowledgements}

\bibliographystyle{aa}    
\bibliography{aa_redsg}


%
\appendix

\section{Remarks on individual stars}

\begin{enumerate}

\item
\object{Sun}:

The 675.7\pun{nm} sulphur triplet is not blended and it is
well reproduced in comparison with synthetic spectra,
when using the atomic data reported in Table \ref{atomicdata}.
We do not consider the sulphur triplet at 674.3\pun{nm} 
whose shape in the solar observed spectra is not well reproduced,
perhaps due to blending by a CN line.
The 674.8\pun{nm} sulphur triplet is also discarded
because of a blend with vanadium, calcium, iron and titanium whose
atomic data are not well known.

The two sulphur lines of Mult.~6 in the range 870\pun{nm} are both well reproduced
by synthetic spectra.
The lines are situated very close to each other, so in the 1D 
analysis we fitted both simultaneously.
In the 3D analysis we concentrated on the 869.4\pun{nm} line.
The 869.3\pun{nm} line is blended with molecules
(CN and C$_2$), and it is weaker than the 879.4\pun{nm} line.
We computed the EW of the contribution of molecules in the
range of this line to
derive the sulphur abundance from this line.  

We do not consider in this work 
the Mult.~1 (920\pun{nm}) sulphur lines, because they are
affected by
telluric lines. The 922.8\pun{nm} line is the cleanest among the
three lines, but it lies in the wing of the Paschen $\zeta$ H-line, so the
abundance analysis is not free from possible systematic errors.

\item
\object{Procyon}:

\citet{2002ApJ...567..544A} and later \citet{2005ApJ...633..424A}
investigated Procyon's limb darkening with 3D models. Both groups found that 3D
models predict a smaller degree of limb darkening than 1D models.
Aufdenberg et al.  showed that an approximate overshooting introduced in 1D
Phoenix or ATLAS models can largely eliminate these differences.
The case of Procyon is fortunate because it is a well studied visual binary system.
The mass of the primary is well determined (see \citealt{2006AJ....131.1015G} 
and  \citealt{2000AJ....119.2428G} for the most 
recent astrometric studies of this star).

The angular diameter has been measured directly  by 
\citet{1967MNRAS.137..393H,1974MNRAS.167..121H} and, more recently,  by  
\citet{1988ApJ...327..905S,1991AJ....101.2207M,1998AA...339..858D,2001AJ....122.2707N,2003AJ....126.2502M,2004AA...413..251K}.

The distance from Hipparcos and orbital motion studies are
reported in \citet{2006AJ....131.1015G}.

The surface gravity of this star is well known
from the orbital data and diameter measurements:  \glog = 4.0 with an error less than 0.1,
in agreement with \citet{2002ApJ...567..544A}, who
obtained \glog = $3.96\pm 0.02$.

The effective temperature depends on the method used to derive it.
Critical reviews of values obtained by various authors with 
different methods are summarised in \citet{1985AAS...59..403S} and 
in \citet{1996PASJ...48..601K}.
The \teff values range from 6\,400\pun{K} (the lowest value derived 
from the continuum and IR fluxes) to more than 6\,800\pun{K} (from the ionisation
balance of Fe). 
By excluding the high values derived from line spectrum analysis,
the highest \teff value depends on the
value of the mixing-length adopted in the models; a lower 
mixing-length corresponds to a lower convective flux so to a higher 
temperature gradient at the bottom of the the atmosphere
\citep{1982AA...113..135K}.
A synthetic spectrum at an effective temperature of 6\,500\pun{K} 
with $\mlp=0.5$
agrees with Balmer line profiles \citep{1994AA...285..585F}.

We give more weight to the value derived 
from the flux distribution, adopt 6\,500\pun{K} and ascribe  the higher
values required to fit the ionisation balance to the inaccuracy of the structure 
of the atmosphere adopted in 1D models. These models require a  higher \teff to
describe the outer atmospheric layers where the lines are formed. 

A further proof of the inadequacy of the structure of the 1D models is 
given by the extensive discussion on the derivation of the atmospheric 
parameters by \citet{2005AJ....129.1063L} based on the choice of the 
model that better fits the spectroscopic data; they obtain
\teff = 6850\pun{K}, \glog~= 4.55,  $\xi_{\rm micro}$ = 2.4\kms.

\begin{table*}
\begin{center}
\caption{Procyon: stellar parameters with reference.
\label{parprocyon}}
\begin{tabular}{llllll}
\hline
\noalign{\smallskip}
 \teff & \glog & [Fe/H] & $\xi_{\rm micro}$ & \vsini &  Reference\\
   K   &       & \kms & \kms \\
\noalign{\smallskip}
\hline
\noalign{\smallskip}
&&&                 & $2.8$        & \citet{1981ApJ...251..152G}\\
 $6500\pm 80$ & $4.0\pm 0.1$ & &$2.1\pm 0.3$     & $< 4.5$      & \citet{1985AAS...59..403S}\\
 6605 & 4.13           & --0.06 & $2.23$           &              & \citet{1996AA...314..191G}\\
 6500 & 4.00           & & $1.4$            & $3.3$        & \citet{1996PASJ...48..511T}\\
 6470 & $4.01\pm 0.10$ & $-0.01\pm 0.07$ & $1.91\pm 0.20$   & $2.6\pm 1.0$ & \citet{1998AA...338..161F}\\
 6640 & 4.13           &&$1.8$            & $6.6$        & \citet{1998PASJ...50..509T}\\
$6530\pm  50$ & $3.96\pm 0.02$ &&$2.2$            & $3.16\pm 0.5$& \citet{2002ApJ...567..544A}\\
$6543\pm  84$ & $3.975\pm 0.013$ & & & & \citet{2005ApJ...633..424A}\\
\noalign{\smallskip}
\hline
\end{tabular}
\end{center}
\end{table*}

A selection of stellar parameters from the literature is given in Table~\ref{parprocyon}.

\item
\object{HD 33256}:

HD~33256 is slightly cooler than Procyon (F5 IV-V)
in spite of the earlier
spectral type given in Simbad and the Bright Star Catalog (BSC), F2V. 
It is slightly metal deficient and 
so has weaker lines 
than a  solar abundance star. This is at the origin of its earlier 
than Procyon's spectral type.
In fact the new accurate classification by \citet{2003AJ....126.2048G}
is F5.5V (kF4, mF2).

It is not far from the Galactic plane,
in fact its Galactic coordinates are: 208.83, --24.83.
It is a thin disk star according to \citet{2003AA...410..527B}.
The S abundance has been measured by \citet{2002ApJ...573..614T}.
These authors adopted [Fe/H] from
\citet{1993AA...275..101E}, while they derived temperature
and gravity (see Table \ref{hd33256par}).
The LTE derived S abundance, 
based on the 8\,693\pun{nm} and 8\,694\pun{nm} lines, is 7.29.
A selection of stellar parameters is listed in Table \ref{hd33256par}.

\begin{table*}
\begin{center}
\caption{Stellar parameters of HD 33256
\label{hd33256par}}
\begin{tabular}{llllll}
\hline
\\
\teff & log g & [Fe/H] & $\xi_{\rm micro}$ & \vsini &  Ref\\
K    &       &        & \kms & \kms \\
\\
\hline
\\
6\,550 & 4.09  & --0.34 & 1.4  & & \citealt{1981AA....97..145N}\\
6\,270 & 4.0 &    &      & & \citealt{1981ApJ...250..262C}\\
6\,270 & 4.0 & --0.26 & 1.0     & & \citealt{1985ApJ...290..289T}\\
6\,440 & 4.05  & --0.30 &      & & \citealt{1991MNRAS.253..610L}\\
6\,300 & 3.60  & --0.45 & 1.40 & & \citealt{1991AA...244..425Z}\\
6\,400 & 3.95 & --0.336 & & 0 &\citealt{1992ApJ...387..170F}\\
6\,442 & 4.05  & --0.30 &      & & \citealt{1993AA...275..101E}\\
6\,386 & 4.10  & --0.30 & 2.10 & & \citealt{1995AJ....109..383K}\\
6\,385 & 4.10  & --0.30 &      & & \citealt{2001AA...371..943C}\\
6\,440 & 3.99  & --0.30 & 2.3  & & \citealt{2002ApJ...573..614T}\\
6\,411 & 3.87  & --0.30 & 1.5  & & \citealt{2003AJ....126.2048G}\\
6\,427 & 4.04  & --0.30 & 1.90 & & \citealt{2003AA...410..527B}\\
     &       &        &      & 9.7 & \citealt{2003AA...398..647R}\\
\\
\\
\hline
\end{tabular}
\\
\end{center}
\end{table*}

\item
\object{HD 25069}:

No detailed abundance analysis has been done for this star, but many
measures of its radial velocity exist up to 2005.
\citet{2005ApJS..159..141V}
made an extensive study of the spectroscopic 
properties of cool stars. From one Keck spectrum
they derived (with Kurucz models):
\teff = 4\,994\pun{K}, \glog~ = 3.53, [M/H] = 0.10,
\vsini = 3.3 \,\kms, RV = 39.2\,\kms.
Abundances of Na, Si, Ti, Fe, Ni are also given.
The star seems to be cooler than spectral type G9, as found in Simbad data base.
We keep the stellar parameters from \citet{2005ApJS..159..141V}.

It is remarkable that, although for such a cool star the 
H$\alpha$ profile is very little sensitive to effective temperature,
from the fitting of the wings of
H$\alpha$ we obtain: \teff = 4\,726\pun{K}, which is in very good
agreement with the temperature derived from the B-V colour:
B-V=1.00 implies \teff = 4\,700\pun{K}.

\item
\object{$\epsilon$ Eri}:

\begin{table*}
\begin{center}
\caption{$\epsilon$ Eri (HD~22049): stellar parameters with reference.
\label{hd22049par}}
\begin{tabular}{lrrrrl}
\hline
\noalign{\smallskip}
\teff & \glog & [Fe/H] & $\xi_{\rm micro}$ & \vsini &  Reference\\
K    &       &        & \kms & \kms \\
\noalign{\smallskip}
\hline
\noalign{\smallskip}
5\,180 & 4.75  & --0.09 & 1.25 &      & \citet{1993ApJ...412..797D}\\
5\,076 & 4.50  &   0.05 &      &      & \citet{1996AAS..117..227A}\\
       & 4.84  &        &      & 2.1  & \citet{2000ApJ...528..885A}\\
5\,135 & 4.70  & --0.07 & 1.14 &      & \citet{2001AA...373.1019S}\\
5\,117 &       &        &      & 3    & \citet{2004AA...418..989N}\\
5\,073 & 4.43  & --0.13 & 1.05 &      & \citet{2004AA...415.1153S}\\
4\,992 &       &        &      &      & \citet{2005ApJ...626..465R}\\
5\,177 & 4.72  &   0.06 & 0.62 &      & \citet{2005PASJ...57...27T}\\
5\,200 & 4.50  & --0.04 & 0.70 &      & \citet{2005AJ....129.1063L}\\
\noalign{\smallskip}
\hline
\end{tabular}
\end{center}
\end{table*}

$\epsilon$ Eri (HD~22049) is a much studied star (795 papers since 1983).
It is a variable star of BY~Dra type
(BY~Dra stars are flare stars with cool spots, which cause
photometric variations during the rotation of the star).
It has two suspected planets: 
$\epsilon$ Eri b \citep{1988ApJ...331..902C},
$\epsilon$ Eri c \citep{2002ApJ...578L.149Q}.
\citet{2006AJ....132.2206B} determined the mass of $\epsilon$ Eri b
from HST and ground-based astrometric and RV data,
modelled its orbit, confirmed the
existence of this companion and discuss the existence of the possible
tertiary invoked by \citet{2002ApJ...578L.149Q} and
\citet{2000ApJ...537L.147O}.
The star has a high level of magnetic activity inferred from
chromospheric activity consistent with a relatively young age, 
less than 1\pun{Gyr}.
Observational and theoretical searches for the signature of
planetary/brown dwarf objects in the structure of the dust disk around this
star are underway by \citet{2006AJ....132.2206B}.
A selection of stellar parameters is listed in Table \ref{hd22049par}.

\end{enumerate}


\end{document}